\documentclass[%
 reprint,
superscriptaddress,
 amsmath,amssymb,
 aps,
prb,
floatfix,
]{revtex4-2}

\usepackage{graphicx}% Include figure files
\usepackage{dcolumn}% Align table columns on decimal point
\usepackage{bm}% bold math
\usepackage{color}
\usepackage{ulem}
\usepackage[T1]{fontenc}
\usepackage{enumitem}

\usepackage{chemscheme}

\makeatletter
\@ifpackageloaded{floatrow}{%
  \floatsetup[table]{capposition=top}%
}{%
  \@ifpackageloaded{float}{%
    \floatstyle{plaintop}\restylefloat{table}%
  }{}%
}
\makeatother

\begin{document}

%\preprint{APS/123-QED}

\title{Doubling the size of quantum selected configuration interaction based on seniority-zero space and its application to QC-QSCI-AFQMC}% Force line breaks with \\

\newcommand{\affilA}{%
Center for Quantum Information and Quantum Biology,
The University of Osaka, 1-2 Machikaneyama, Toyonaka, Osaka 560-0043, Japan
}%
\newcommand{\affilB}{%
Graduate School of Engineering Science, The University of Osaka, 1-3 Machikaneyama, Toyonaka, Osaka 560-8531, Japan
}%
\newcommand{\affilC}{Technology Strategy Center, TOPPAN Digital Inc., 1-3-3 Suido, Bunkyo-ku, Tokyo 112-8531, Japan}
\newcommand{\affilD}{Technical Research Institute, TOPPAN Holdings Inc., 4-2-3 Takanodai-minami, Sugito-machi, Kita-katsushika-gun, Saitama 345-8508, Japan}

\author{Yuichiro~Yoshida}
\email{yoshida.yuichiro.qiqb@osaka-u.ac.jp}
\affiliation{\affilA}
\author{Takuma~Murokoshi}
\affiliation{\affilA}
\author{Rika~Nakagawa}
\affiliation{\affilC}
\author{Chihiro~Mori}
\affiliation{\affilC}
\author{Yuta~Katayama}
\affiliation{\affilC}
\author{Naoya~Kuroda}
\affiliation{\affilA}
\author{Shigeki~Furukawa}
\affiliation{\affilD}
\author{Hanae~Tagami}
\affiliation{\affilD}
\author{Wataru~Mizukami}
\email{mizukami.wataru.qiqb@osaka-u.ac.jp}
\affiliation{\affilA}%

\date{\today}% It is always \today, today,

\begin{abstract}
We propose doubly occupied configuration interaction-quantum selected configuration interaction (DOCI-QSCI), which samples from the seniority-zero space. 
While the use of this space effectively doubles the qubit budget, equaling the number of spatial orbitals, this sector restriction can compromise quantitative accuracy.
To compensate for this, we expand sampled bitstrings via their Cartesian product into a larger space that includes seniority-breaking determinants.
The resulting wave function is also proposed using the trial state in phaseless auxiliary-field quantum Monte Carlo (ph-AFQMC) to recover dynamical correlations across the full orbital space (DOCI-QSCI-AFQMC).
We evaluate the proposed methods on the H$_6$ chain, N$_2$ dissociation, and the addition of singlet O$_2$ to a BODIPY dye.
For the H$_6$ chain, DOCI-QSCI-AFQMC reproduces the accuracy of the level of the complete-active-space counterpart with the quantum device ibm\_kobe.
For N$_2$ and BODIPY–O$_2$, with (14e, 28o) and up to (20e, 20o) active spaces, it yields reasonable results, whereas single-reference CCSD(T) fails qualitatively.
These results demonstrate that the DOCI-QSCI doubles the orbital space accessible to conventional QSCI and subsequent ph-AFQMC post-processing delivers reasonably high accuracy.
\end{abstract}

\maketitle

\section{Introduction \label{sec:intro}}

Quantum chemistry on quantum computers has advanced rapidly over the past decade~\cite{Peruzzo2014variational,Kandala2017hardware,Huggins2022unbiasing,Kanno2023quantum,Robledo-Moreno2025chemistry}.
While the rapid development of quantum hardware has been crucial to this progress, improvements in quantum algorithms---particularly refinements to quantum--classical hybrid workflows---have been equally important.
Although current quantum-chemical calculations on quantum computers are typically approximate, they are now approaching problem sizes for which finding classically exact solutions is no longer feasible on conventional (classical) computers.
They are beginning to compete with state-of-the-art classical methods, such as coupled cluster (CC) and density matrix renormalization group calculations. In light of these developments, performing practically relevant calculations is becoming realistic on today's quantum hardware.

Quantum-selected configuration interaction (QSCI)~\cite{Kanno2023quantum} is one such algorithm that has shown notable success.
A prominent demonstration is the calculation of Fe--S clusters using up to 77 qubits~\cite{Robledo-Moreno2025chemistry,Shirakawa2025closed}, enabled by a powerful error mitigation protocol that supports quantum--classical hybrid computation at this scale.
This corresponds to a (54e,36o) active space calculation, which is beyond what can be readily treated with classical quantum-chemistry methods, highlighting the substantial potential of QSCI.

Despite this success, further improvements are still needed to make more effective use of QSCI~\cite{Nakagawa2024ADAPT,Sugisaki2025Hamiltonian,Mikkelsen2024Quantum,Chen2026neural}.
In particular, the standard fermion-to-qubit mapping, in which the number of qubits scales one-to-one with the number of spin orbitals, drastically increases the size of the many-qubit Hilbert space and the number of possible entangling operations, pushing circuit depths well beyond what is feasible.
Constant-factor reductions achieved through symmetry-based qubit tapering~\cite{Bravyi2017tapering,Seria2020reducing} may not be sufficient.
Aggressive compression that disregards the structure and symmetries of the Hilbert space risks undermining the design principles of quantum circuits.

A natural way to partition Hilbert space is by the seniority number, a long-standing concept in quantum chemistry~\cite{Bytautas2011seniority,Bytautas2015seniority} and in nuclear and condensed-matter physics.
The seniority number is defined as the number of unpaired electrons in a Slater determinant, and the seniority-zero subspace comprises electronic configurations with no unpaired electrons.
By retaining only paired electrons, the seniority-zero space captures the pair correlations essential for static correlation and is known to describe static correlation effectively.
Moreover, doubly occupied configuration interaction (DOCI)~\cite{Allen1962electron,Smith1965natural,Weinhold1967reduced,Veillard1967complete,Cook1975doubly,Fantucci1977direct,Couty1997generalized,Kollmar2003new}---i.e., configuration interaction (CI) restricted to the seniority-zero space---involves far fewer configurations than full CI, making it more tractable as the number of orbitals increases.

Conversely, it is known that the seniority-zero space captures only a limited subset of electronic correlation.
Achieving fully quantitative accuracy ultimately requires enabling seniority breaking, and the seniority-zero space provides a practical foundation for this.

The feasibility of utilizing the seniority-zero space for quantum computation has been explored~\cite{Elfving2011simulating,Kottmann2022optimized,Kim2024variational,Zhao2023orbital,Zhao2024enhancing,O’Brien2023purification,Khamoshi2021correlating,Khamoshi2023AGP,Khamoshi2023state}.
The primary advantage is that the number of qubits can be equal to the number of spatial orbitals rather than the number of spin orbitals, effectively halving the qubit count.
Consequently, this also enables compact and readily implementable quantum-circuit designs.
Variational algorithms, circuit implementations, and state preparation for QPE have been studied.
Therefore, utilizing the seniority-zero space in quantum computing is a promising research direction, offering avenues for further algorithmic improvements.

In this paper, we propose a QSCI method that samples from the seniority-zero space. Moreover, we incorporate classical postprocessing via a ph-AFQMC calculation to elevate QSCI to highly accurate quantum-chemical calculations.
As one approach to improving accuracy through seniority breaking, we expand each sampled result into two spin-resolved configurations and treat them in a Cartesian product space.
The resulting QSCI wave function is used as the trial wave function for the ph-AFQMC calculation to recover electronic correlation---primarily dynamical correlation---across the full orbital space.
The proposed method is tested on benchmark molecular systems, such as a hydrogen chain and N$_2$.
The present method is also applied to the oxygen addition reaction of a BODIPY dye, demonstrating its applicability to a practical molecular system.

%Yoshida2025QSCIAFQMC,Danilov2025enhancing,Erhart2025coupled,Shirai2025enhancing,

  %I believe leaving the sections in separate files is more organized, change it if you desire 
\section{Computational methods \label{sec:methods}}

\subsection{Quantum-selected configuration interaction and its extension}

\subsubsection{Basic formalism of QSCI}

This section provides a brief review of the basic formalism of QSCI~\cite{Kanno2023quantum}.
The QSCI method constructs the effective Hamiltonian 
\begin{align}
    \hat{H}_\text{eff} = \hat{P}_S\hat{H}\hat{P}_S
\end{align}
and then obtains its eigenstates.
Here, $\hat{H}$ is the electronic structure Hamiltonian, and in the case of active-space QSCI, it corresponds to the active-space Hamiltonian.
This effective Hamiltonian is constructed based on the sampling results from a quantum computer. It defines a subspace $S$ spanned by the computational basis states associated with the quantum state of interest.
The operator $\hat{P}_S$ is the projector onto $S$, and can be expressed as 
\begin{align}
    \hat{P}_S = \sum_{i=1}^R | \Phi_i \rangle \langle \Phi_i |
\end{align}
using the computational basis states $\{|\Phi_i\rangle\}$ spanning the subspace $S$.

The eigenstates of the effective Hamiltonian $\hat{H}_\text{eff}$ are obtained by solving the following equation:
\begin{align}
    \hat{H}_\text{eff} | \Psi \rangle = E | \Psi \rangle
\end{align}
on a classical computer, where $| \Psi \rangle$ and $E$ are the eigenvector and the corresponding eigenvalue, respectively.

\subsubsection{Use of Cartesian product of bitstrings}

In this section, we review an approach that accelerates sampling and facilitates the construction of spin-adapted wave functions~\cite{Yoshida2025QSCIAFQMC,Robledo-Moreno2025chemistry}.
Because a bitstring $\Phi_i$ corresponds to a Slater determinant, it can be written as the concatenation of the $\alpha$- and $\beta$-spin parts:
\begin{align}
    \Phi_i &= n_{1\alpha,i}n_{2\alpha,i} \cdots n_{N\alpha,i}n_{1\beta,i}n_{2\beta,i} \cdots n_{N\beta,i} \\
           &:= \Phi_i^{(\alpha)}\Phi_i^{(\beta)},
\end{align}
where $n_{k\sigma,i}$ is the occupation number of the $k\sigma$ spin-orbital (the $\sigma$-spin component of the $k$th spatial orbital) in the $i$th configuration.
Here, $N$ is the number of spatial orbitals.

When forming determinants by concatenating $\alpha$ and $\beta$ bitstrings, taking the Cartesian product of the two pools, 
\begin{align}
    \{\tilde{\Phi}_k\} = \{ \Phi_i^{(\alpha)}\Phi_j^{(\beta)}, ^\forall i, j \leq R\} \label{eq:Cart}
\end{align}
can generate a set that is larger than the originally sampled set.

\subsubsection{DOCI-QSCI}

In this work, we further improve the efficiency of handling the electronic configurations introduced above.
We perform quantum computation and measurement within the seniority-zero subspace and obtain a set of bitstrings $\{ \Phi_i^{(0)}\}$.
Here, the superscript ``0'' denotes the seniority-zero (pair-occupation) representation.

We define the $\alpha$- and $\beta$-string pools from the same sampled set as
\begin{align}
    \{\Phi_i^{(0)}\} = \{\Phi_i^{(\alpha)}\} = \{\Phi_i^{(\beta)}\}, \quad ^\forall i \leq R,
\end{align}
and then generate spinful determinants by combining the two pools using Eq.~(\ref{eq:Cart}).

Thus, although the sampling is performed in the DOCI (seniority-zero) representation, the subspace used to construct the effective Hamiltonian in QSCI is not restricted to seniority zero.
This procedure reduces the cost of determinant selection by sampling in an $N$-orbital representation, while the effective Hamiltonian is constructed in the Cartesian product space guided by the seniority-zero sampling.

\subsection{DOCI-QSCI-AFQMC workflow}

In this work, we incorporate DOCI-QSCI into the QC-QSCI-AFQMC workflow (hereafter, QSCI-AFQMC).
In QSCI-AFQMC~\cite{Yoshida2025QSCIAFQMC,Danilov2025enhancing}, the QSCI wave function is used as the trial wave function for phaseless (ph) AFQMC~\cite{Zhang2003quantum} to recover the missing electron correlation across the full orbital space.
Likewise, in DOCI-QSCI-AFQMC, the DOCI-QSCI wave function is used as the trial wave function for ph-AFQMC.

\begin{figure*}[ht]
    \centering
    \includegraphics[width=1.0\linewidth, bb=0 0 1418 441]{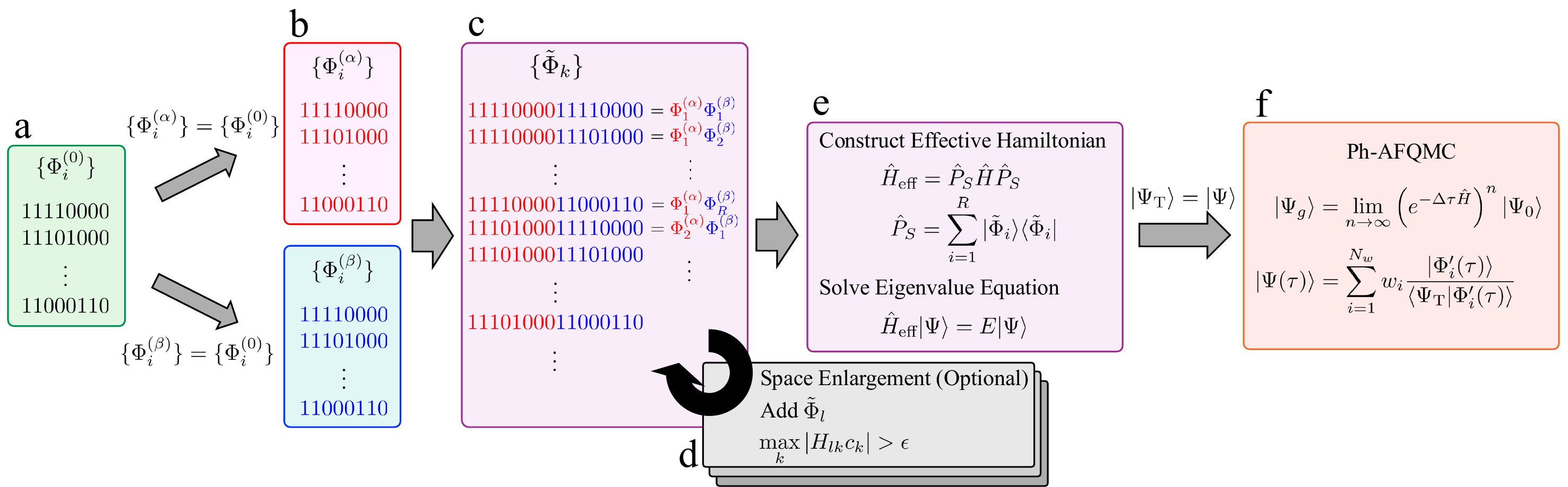}
    \caption{DOCI-QSCI-AFQMC workflow.}
    \label{fig:flow}
\end{figure*}

Figure~\ref{fig:flow} shows the workflow of DOCI-QSCI-AFQMC.
The main steps are as follows.
\begin{enumerate}[label=\alph*.]
    \item Sample from the seniority-zero space and obtain bitstrings of length $N$.
    \item Generate the $\alpha$- and $\beta$-string pools from the sampled seniority-zero set $\{\Phi_i^{(0)}\}$.
    \item Take the Cartesian product of $\{\Phi_i^{(\alpha)}\}$ and $\{\Phi_i^{(\beta)}\}$.
    \item Expand the selected space spanned by $\{\tilde{\Phi}_k\}$ using these determinants as seeds (optional).
    \item Construct and diagonalize the effective Hamiltonian $\hat{H}_{\text{eff}}$.
    \item Perform a ph-AFQMC calculation using the DOCI-QSCI trial wave function.
\end{enumerate}

If the sampling in the seniority-zero space is terminated after selecting $R$ determinants, the number of determinants increases from $R$ to $R^2$ when all $\alpha$--$\beta$ combinations are formed.
As a result, the selected subspace (and thus $\hat{H}_{\text{eff}}$) includes not only seniority-zero determinants but also broken-pair determinants with seniority $>0$, which can improve quantitative accuracy.
In ph-AFQMC, the quality of the trial wave function $|\Psi_\text{T}\rangle$ is crucial for Monte Carlo sampling; therefore, ensuring a high-quality $|\Psi_\text{T}\rangle$ is essential.

We also implement an optional expansion of the selected-determinant space following the conventional selected-CI approach.
An external candidate determinant $l$ is added to the set $\{\tilde{\Phi}_k\}$ if its importance score $\max_k | H_{lk}c_k|$ exceeds the cutoff $\epsilon$, where $k$ runs over the current selected determinants. Here, $H_{lk} = \langle \tilde{\Phi}_l |\hat{H}| \tilde{\Phi}_k \rangle$ is the $(l,k)$ element of the Hamiltonian, and $c_k$ is the expansion coefficient of determinant $k$ in the wave function.

\section{Computational details} \label{sec:comput_details}

Self-consistent field (SCF) calculations for generating molecular integrals, as well as conventional quantum-chemical calculations including CC singles and doubles (CCSD) and CCSD with perturbative triples (CCSD(T)), were performed using PySCF~\cite{PySCF,PySCF2}.
We used the cc-pVTZ, cc-pVQZ, and 6-31G(d,p) basis sets for the H$_6$ chain, N$_2$, and BODIPY--O$_2$ complex, respectively.
Reference heat-bath configuration interaction (HCI) calculations were performed using PyCI~\cite{PyCI}.

Quantum computations were performed using the Qiskit library v2.2.3~\cite{qiskit2024}.
We employed the spinless unitary cluster Jastrow (UCJ) ansatz with a single repetition.
The Jastrow interactions were restricted to adjacent orbitals (qubits); i.e., in a local UCJ manner~\cite{Motta2023bridging}.
The circuit parameters were fixed to the active-space amplitudes obtained from frozen-core CCSD calculations.
We used an IBM Quantum device \texttt{ibm\_kobe} and a noiseless simulator. When using the real quantum device, sampling results that do not preserve the number of electrons are physically meaningless and are discarded.

We used the space sampled by QSCI and the expanded space in an HCI fashion. These two strategies were performed via the \texttt{kernel\_fixed\_space} and \texttt{kernel\_float\_space} functions in PySCF, respectively. Hereafter, $\epsilon$ is given as \texttt{select\_cutoff} in PySCF.

The ph-AFQMC calculations were conducted using ipie v0.6.2~\cite{Malone2023ipie}.
We employed a time step of 0.005 $E_h^{-1}$, 50 steps, and 3000 blocks, unless otherwise specified.

The reaction path for the addition of singlet oxygen to the BODIPY dye was investigated at the RB3LYP/6-31G(d) level and identified by intrinsic reaction coordinate calculations.
Density functional theory calculations were carried out using Gaussian~16~Rev.~C~\cite{g16}.

\section{Results and discussion} \label{sec:results}

Here, we present the computational results of DOCI-QSCI-AFQMC and discuss their validity. 
First, we present results for the dissociation of the H$_6$ chain and N$_2$, typical molecular systems for strong electron correlation.
Next, we examine the cyclization reaction of a BODIPY molecule with singlet oxygen as an example of a realistic chemical reaction.

Throughout this paper, the name of the wave function theory calculation used for the ph-AFQMC trial wave function is placed before ``AFQMC'' (e.g., QSCI-AFQMC, CASCI-AFQMC).

\subsection{H$_6$ chain}

\begin{figure*}[ht]
    \centering
    \includegraphics[width=1.0\linewidth, bb=0 0 1800 680]{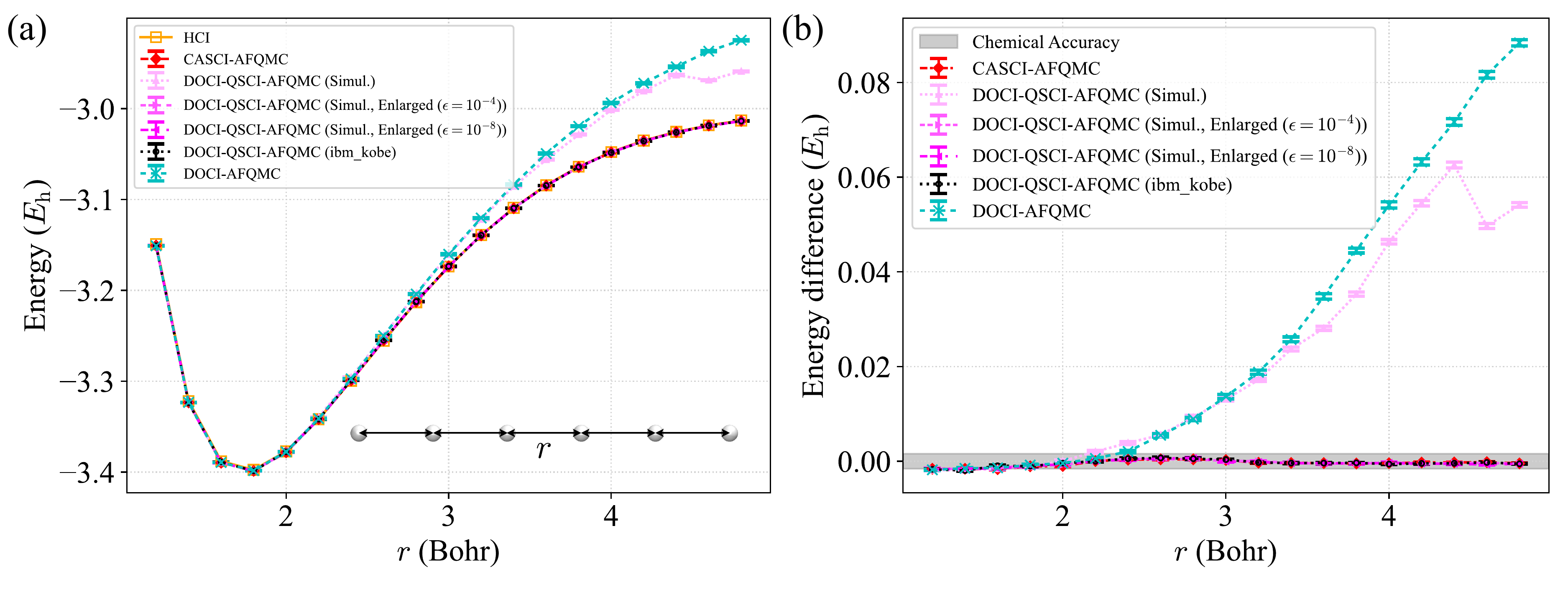}
    \caption{(a) Energy curve and (b) energy difference relative to HCI for the one-dimensional H$_6$ chain. The horizontal axis is the distance between adjacent hydrogen atoms. Simul. denotes results obtained with a classical simulator, and $\epsilon$ is the cutoff for the selected CI space enlargement. Sampling was performed $10^5$ shots at each point.}
    \label{fig:H6}
\end{figure*}

Figure~\ref{fig:H6}(a) and (b) show the energy curves and the energy differences relative to HCI for the one-dimensional H$_6$ chain, including DOCI-QSCI-AFQMC results obtained with a classical simulator and on the quantum computer \texttt{ibm\_kobe}.
For DOCI-QSCI-AFQMC calculations used by the simulator, we also report results in which the sampled subspace is further enlarged using a conventional selected-CI procedure. We used 400 walkers in the ph-AFQMC calculations.

DOCI-QSCI-AFQMC using \texttt{ibm\_kobe} reproduces the CASCI-AFQMC results, indicating that the subspace $S$ used to construct the effective Hamiltonian is equivalent to the complete active space in this case.
Another point to note is that both curves closely match the HCI reference. 
Although the three points to the left of the energy minimum deviate slightly beyond chemical accuracy relative to HCI, the remaining points are within chemical accuracy. 
Both the near-equilibrium region and the strongly correlated dissociation regime are chemically relevant, so achieving high accuracy in these regions is particularly important.

In contrast, DOCI-QSCI-AFQMC using a noiseless simulator does not reproduce the high-accuracy results unless the subspace is enlarged in the selected-CI manner.
This reflects the fact that with a finite number of samples, the probability distribution defined by the ansatz can be heavily skewed toward a small number of bitstrings, such as the Hartree-Fock configuration, leading to insufficient coverage of the relevant determinants.
However, highly accurate energy values can be readily obtained by expanding the space.

The control calculation using a DOCI trial wave function (DOCI-AFQMC) indicates that DOCI alone does not provide quantitatively reliable results for this kind of strongly correlated system, which is consistent with a previous study~\cite{Yoshida2025auxiliary}.
DOCI-AFQMC yields a larger energy error than DOCI-QSCI-AFQMC, beyond the statistical uncertainty of quantum Monte Carlo sampling.
This highlights the difference between the seniority-zero space and the Cartesian product space of bitstrings guided by seniority-zero sampling.

\subsection{Potential energy curve for N$_2$}

\begin{figure}[ht]
    \centering
    \includegraphics[width=1.0\linewidth, bb=0 0 567 572]{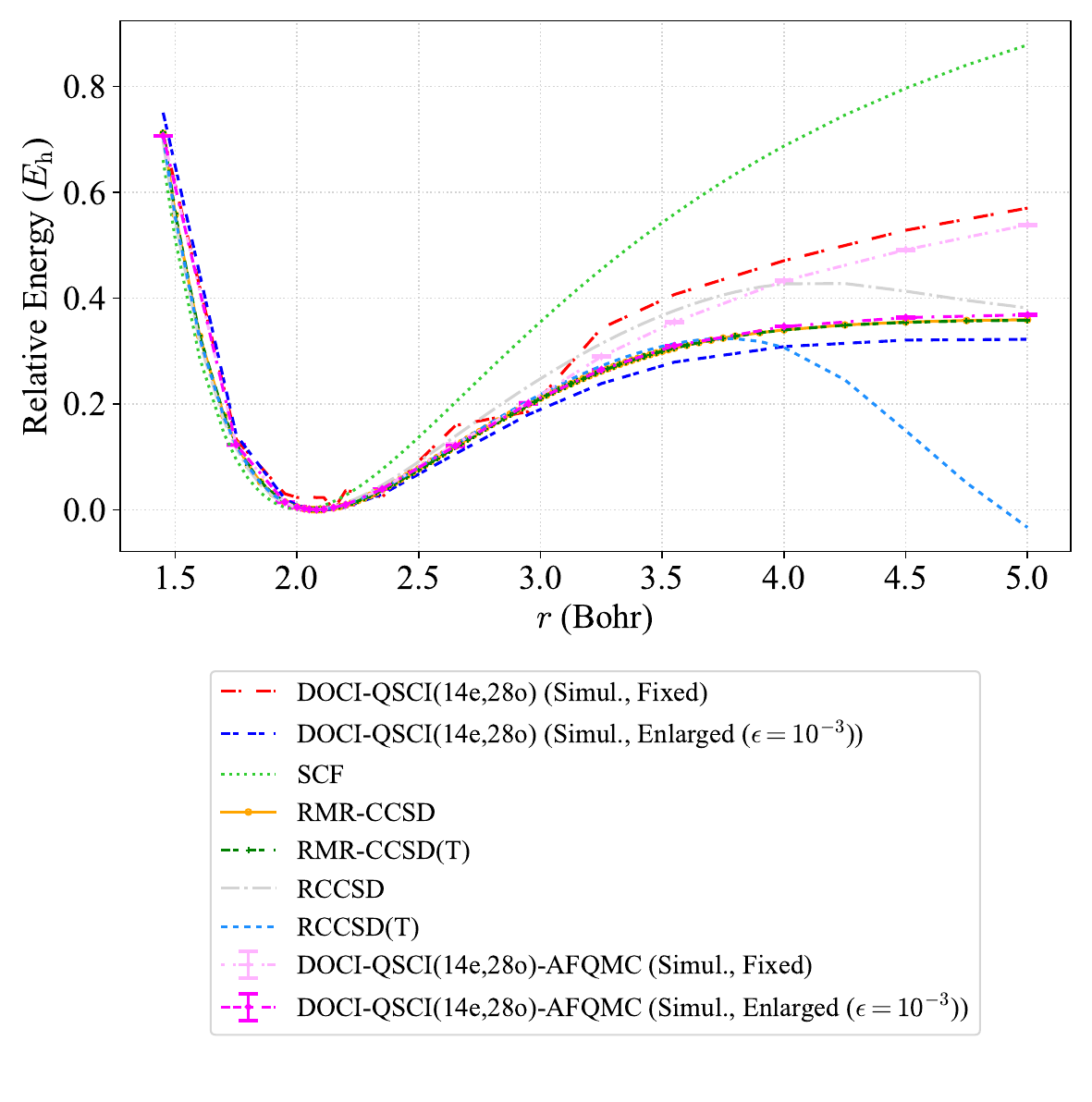}
    \caption{Relative energy curves of N$_2$ in Hartree. Sampling was performed $10^5$ shots at each point. For ph-AFQMC calculations, we used 640 walkers. RMR-CCSD and RMR-CCSD(T) data were taken from Ref.~\cite{Li2008full}.}
    \label{fig:N2}
\end{figure}

Figure~\ref{fig:N2} shows the relative energy curves for N$_2$.
N$_2$ dissociation is often used as a benchmark for strongly correlated systems~\cite{Peterson1993Benchmark,Leszczyk2022assessing,Gdanitz1998accurately,Erhart2024coupled,Khinevich2025enhancing}.
For each method, the relative energies are based on the minimum point in the equilibrium region.
Reduced multireference coupled cluster (RMR-CC) results were obtained from Ref.~\cite{Li2008full}.
RMR-CC is an externally corrected CC approach that serves as a high-accuracy reference in conventional computations~\cite{Li2008full,vspirko2011potential}.

For DOCI-QSCI-AFQMC and DOCI-QSCI, we present two variants: (i) a fixed subspace constructed from sampling (Fixed) and (ii) an enlarged subspace obtained by further expanding the space using a conventional selected-CI procedure (Enlarged).

The $10^5$ shots for each point are insufficient; consequently, DOCI-QSCI with a fixed subspace does not reproduce the dissociation curve quantitatively.
Similarly, DOCI-QSCI-AFQMC, which uses the corresponding DOCI-QSCI trial wave function, also fails to describe the dissociation accurately.
This indicates that the ph-AFQMC postprocessing inherits the deficiency in static correlation present in the trial state.

In contrast, DOCI-QSCI with space enlargement captures the dissociation qualitatively, mitigating finite-sampling effects.
The resulting DOCI-QSCI-AFQMC agrees very well with RMR-CCSD and RMR-CCSD(T).

In contrast, RCCSD and RCCSD(T) yield qualitatively incorrect energy curves. This well-known breakdown of single-reference methods underscores the need for multireference treatments.

\subsection{Reaction between BODIPY dye and singlet oxygen}

In this section, we apply the presented method to the addition of singlet oxygen to a BODIPY molecule.
The addition of singlet oxygen to conjugated systems has attracted interest in modern high-level quantum chemical calculations~\cite{Winslow2024spin}.
To describe configuration interactions associated with bond formation and cleavage, as well as the stabilization of conjugated systems, more orbitals must be included in the active space, which increases the computational cost.
BODIPY dyes are among the most prominent conjugated systems in modern chemistry, and understanding their decomposition pathways is crucial for extending their lifespan~\cite{Mula2008design, Jagtap2013change, Rybczynski2021photochemical}.
Here, we explore the feasibility of treating large active spaces using quantum computing.

The reaction scheme is presented in Scheme~\ref{sch:reaction}.
This scheme is identical to those in previous studies~\cite{Mula2008design,Jagtap2013change}; the chemical structure differs only in the substituents (i.e., the specific simple derivatives considered).

\begin{scheme}[ht]
  \centering
  \includegraphics[width=1.0\linewidth, bb=0 0 487 171]{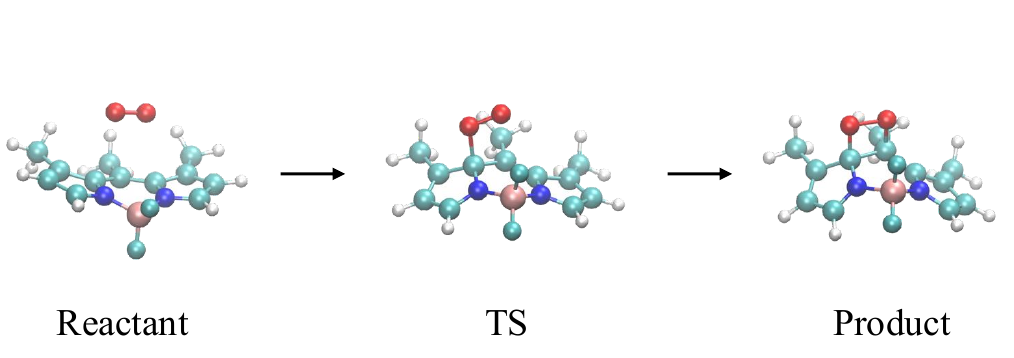}
  \caption{Reaction mechanism for the addition of singlet oxygen to a BODIPY dye. Molecular graphics were created using VMD~\cite{humphrey1996vmd}.}
  \label{sch:reaction}
\end{scheme}

\begin{figure}[ht]
    \centering
    \includegraphics[width=0.9\linewidth, bb=0 0 388 405]{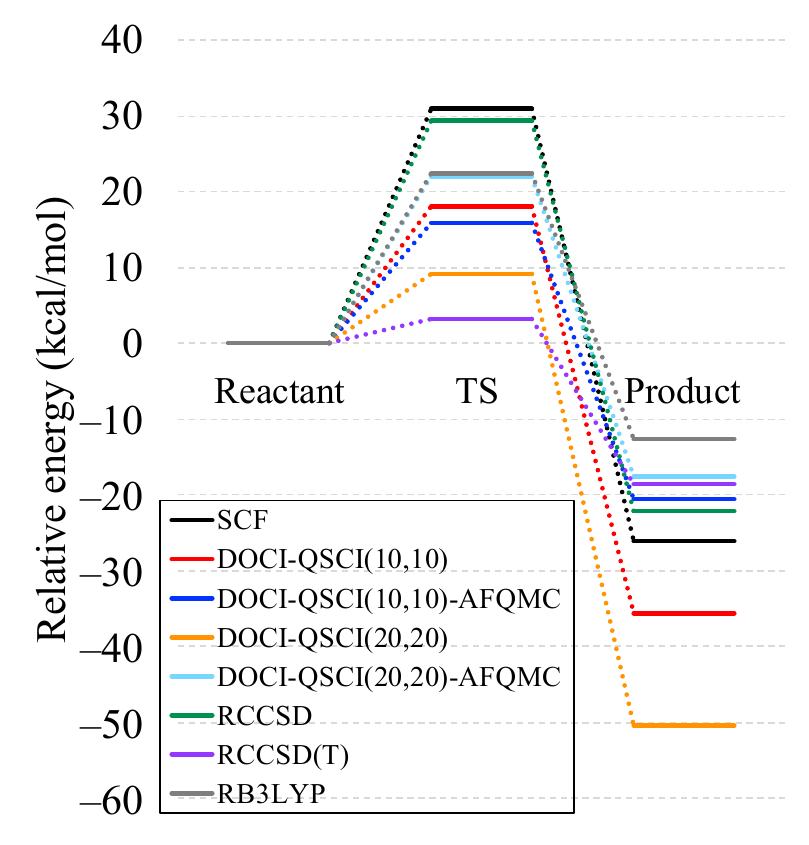}
    \caption{Energy profile of the reaction between BODIPY and O$_2$.}
    \label{fig:BODIPY}
\end{figure}

The activation and reaction energies $(E_a, E_r)$ are shown in Figure~\ref{fig:BODIPY} and Table~\ref{tab:BODIPY}.
Density fitting was used for calculations other than RB3LYP.
DOCI-QSCI and DOCI-QSCI-AFQMC employ natural orbitals obtained from a CISD calculation; increasing the size of active space is therefore expected to improve the accuracy of the resulting energies.
The number of shots for sampling is $10^6$.
For the ph-AFQMC calculations, we used 512 walkers and 1000 blocks.

\begin{table}[htbp] % 260206_BODIPYまとめ.xlsx
    \centering
    \caption{Activation and reaction energies of the reaction between oxygen and BODIPY in kcal/mol.}
    \label{tab:BODIPY}
    \begin{tabular}{l rrr} \hline \hline
      Method          & $E_a$ & $E_r$ \\ \hline
      DOCI-QSCI(20e,20o)-AFQMC & $22.1 \pm 2.4$ & $-17.5 \pm 2.4$ \\
      DOCI-QSCI(10e,10o)-AFQMC & $15.9 \pm 2.0$ & $-20.6 \pm 2.1$ \\
      DOCI-QSCI(20e,20o)       &  9.1 & $-$50.3 \\
      DOCI-QSCI(10e,10o)       & 18.1 & $-$35.6 \\
      RCCSD                     & 29.4 & $-$22.1 \\
      RCCSD(T)                  &  3.2 & $-$18.4 \\
      SCF                      & 30.9 & $-$26.1 \\
      RB3LYP                    & 22.5 & $-$12.6 \\
    \hline \hline
    \end{tabular}
\end{table}

The DOCI-QSCI-AFQMC results show less dependence than the corresponding DOCI-QSCI results on the size of the active space.
This suggests that performing ph-AFQMC in the full orbital space reduces the sensitivity of the computed energies to the choice of active space. 
The resulting activation energy is predicted to be similar to RB3LYP.

The large discrepancy between RCCSD and RCCSD(T) indicates that this system is challenging for single-reference methods.
While the CCSD activation energy is approximately 29~kcal/mol, the CCSD(T) value is around 3~kcal/mol. 
The $T_1$ diagnostic values~\cite{Lee1989diagnostic} are 0.012 (Reactant), 0.068 (TS), and 0.012 (Product).
In particular, the value at the TS exceeds 0.02, suggesting that a multireference treatment is required.
The SCF result is also not quantitatively reliable.

\section{Conclusions} \label{sec:conclusions}

We have proposed DOCI-QSCI, an improved variant of QSCI obtained by sampling in the seniority-zero space.
We have also proposed using the DOCI-QSCI wave function as the trial wave function in the QC-QSCI-AFQMC framework~\cite{Yoshida2025QSCIAFQMC,Danilov2025enhancing}; we refer to this approach as DOCI-QSCI-AFQMC.
Working in the seniority-zero space allows the number of qubits to be equal to the number of spatial orbitals rather than the number of spin orbitals, thereby halving the required number of qubits (or, equivalently, doubling the number of spatial orbitals that can be treated for a fixed qubit budget).
Conversely, because the seniority-zero space is smaller than the full Hilbert space, restricting the wave function to this subspace can affect quantitative accuracy.
In this paper, we considered Cartesian-product states constructed from the sampled bitstrings. We evaluated the QSCI energy in the resulting product space, which extends beyond the corresponding seniority-zero subspace.

The proposed methods were tested on several molecular systems.
In the H$_6$ chain benchmark, DOCI-QSCI-AFQMC provided accurate results with real quantum hardware and in simulation.
In some simulations, we used bitstrings obtained from low-shot sampling as seeds and enlarged the space by adding $\alpha/\beta$ strings in a heat-bath CI fashion.
We also confirmed that the DOCI-QSCI trial wave function was of higher quality than the seniority-zero DOCI trial wave function.

In the N$_2$ benchmark, the DOCI-QSCI-AFQMC relative energy curve was in good agreement with those of RMR-CCSD and RMR-CCSD(T).

In the application to the singlet O$_2$ addition reaction to BODIPY, the activation and reaction energies from DOCI-QSCI-AFQMC were reasonable, whereas RCCSD and RCCSD(T), modern standard SR methods, were considered to fail to provide a reasonable evaluation of the activation energy.

Overall, DOCI-QSCI and DOCI-QSCI-AFQMC, which reduce the required number of qubits, are expected to facilitate the scaling up of practical quantum chemical calculations using quantum computers. Conversely, open-shell and high-spin states are not accessible within the seniority-zero space, and further methodological development is required.

Seniority-zero-based DOCI-QSCI paves the way toward scaling up the size of active spaces that are tractable on quantum computers.
It is unlikely that quantum computers will immediately deliver exact solutions to classically intractable problems without prior information, and it is therefore natural to validate the obtained solutions in an exploratory manner.
In general, QSCI has a significant advantage because the resulting energy carries no statistical uncertainty from sampling.
Instead, the error in QSCI manifests as a model error due to insufficient or inaccurate sampling.
This model error corresponds to a mismatch in the modeled subspace $S$, and thus the accuracy of a QSCI solution leaves room for independent verification.
In this sense, QSCI is a reasonable approach for tackling classically intractable problems.

Moreover, seniority-zero-based DOCI-QSCI efficiently halves the required number of qubits, making it easier to scale quantum computations than previous QSCI approaches.
As quantum-chemical calculations on current quantum hardware are reaching practically relevant scales, DOCI-QSCI can, in principle, access problems at roughly twice that scale. We therefore expect it to move beyond the current practical scale.

Solving classically intractable, strongly correlated problems is generally believed to require fault-tolerant quantum computers (FTQCs). Our proposal has the potential to provide a heuristic glimpse into regimes that are expected to become accessible only with FTQCs.

{\it{Note added.---}}
During finalizing this project, we became aware of Ref.~\cite{McFarthing2026Noise} by Petruccione and co-workers, which also proposes a half-qubit QSCI approach. Both works have been carried out in parallel and independently.

%\vspace{4ex}
\section*{Acknowledgements}

This work was supported by MEXT Quantum Leap Flagship Program (MEXTQLEAP) Grant No. JPMXS0120319794, JST COI-NEXT Program Grant No. JPMJPF2014, JST ASPIRE Program Grant No. JPMJAP2319, and New Energy and Industrial Technology Development Organization (NEDO) Grant No.~JPNP20017.
This research was partially supported by the JSPS Grants-in-Aid for Scientific Research (KAKENHI) Grant No. JP23H03819. 
We thank the Supercomputer Center, the Institute for Solid State Physics at the University of Tokyo for the use of the facilities. This work was also conducted using SQUID at the Cybermedia Center at the University of Osaka.

%\nocite{*}

\bibliographystyle{apsrev4-2}
\bibliography{apssamp}% Produces the bibliography via BibTeX.

\end{document}